\documentclass{article}

\usepackage{arxiv}
\usepackage[utf8]{inputenc}
\usepackage[T1]{fontenc}
\usepackage{hyperref}  
\usepackage{url}
\usepackage{booktabs}
\usepackage{amsfonts}
\usepackage{nicefrac}
\usepackage{microtype}
\usepackage{lipsum}
\usepackage{graphicx}
\usepackage{subfig}
\usepackage{float}

\title{EEG multipurpose eye blink detector using convolutional neural network}

\author{
    
  Amanda Ferrari Iaquinta \\
  Department of Electrical Engineering\\
  IFSP - Federal Institute of Education, Science and Technology of São Paulo\\
  Piracicaba - SP - Brazil \\
  \texttt{a.ferrariiaquinta@gmail.com } \\
  
   \And
   
  Ana Carolina de Sousa Silva \\
  Department of Department of Basic Science - FZEA\\
  University of São Paulo - USP \\
  \texttt{University of São Paulo} \\
  
    \And
   
  Aldrumont Ferraz Júnior \\
  Quickium technology\\
  Piracicaba - SP - Brazil \\
  \texttt{ajunior@quickium.com} \\
  
   \And
   
  Jessica Monique de Toledo \\
  IFSP - Federal Institute of Education, Science and Technology of São Paulo\\
  Piracicaba - SP - Brazil \\
  \texttt{jtoledo@quickium.com} \\
  
  \And
  
  Gustavo Voltani von Atzingen \\
  IFSP - Federal Institute of Education, Science and Technology of São Paulo\\
  Rua, Av. Diácono Jair de Oliveira, 1005 \\
  Piracicaba - SP - Brazil \\
  \texttt{gustavo.von@ifsp.edu.br} \\
  
}

\begin{document}
\maketitle

\begin{abstract}
    The electrical signal emitted by the eyes movement produces a very strong artifact on EEG signal due to its close proximity to the sensors and abundance of occurrence. In the context of detecting eye blink artifacts in EEG waveforms for further removal and signal purification, multiple strategies where proposed in the literature.  Most commonly applied methods require the use of a large number of electrodes, complex equipment for sampling and processing data. The goal of this work is to create a reliable and user independent algorithm for detecting and removing eye blink in EEG signals using CNN (convolutional neural network). For training and validation, three sets of public EEG data were used. All three sets contain samples obtained while the recruited subjects performed assigned tasks that included blink voluntarily in specific moments, watch a video and read an article. The model used in this study was able to have an embracing understanding of all the features that distinguish a trivial EEG signal from a signal contaminated with eye blink artifacts without being overfitted by specific features that only occurred in the situations when the signals were registered.
\end{abstract}


\section{Introduction}
Electroencephalography had its origin by the 1920s, Hans Berger, a neuropsychiatrist from Germany, recorded potentials from the scalp of patients with skull defects and, a few years later, with more sensitive equipment from intact subjects. Berger published over 20 reports on the electroencephalogram, a term he introduced in 1929\cite{stone2013early}.

In the 1960s, electroencephalography (EEG) was tied to the laboratory due to equipment and recording requirements. Today the field has moved from simple and artefact-sensitive EEG recording to making real the vision of brain-computer communication. In the last 40 years, direct brain-computer interaction went from simple communication programs to sophisticated BCI-controlled applications.  Generally speaking, a brain-computer interface (BCI) is a device that connects the brain to a computer and decodes in real time a specific, predefined brain activity. This brain activity has to be measured either directly, via the electrical activity of nerve cells, or indirectly \cite{kubler2019history}. 

The most convenient and widely used method for recording brain activity in BCI applications is electroencephalography (EEG). During EEG, brain electrical activity is recorded by placing electrodes on the scalp. This method has high temporal resolution and is safe, easy to use, and affordable \cite{islam2016methods}\cite{ramadan2017brain}. 

Electrodes are placed on specific scalp locations, for example over parietal areas toward the back of the skull to measure activity related to attentional and memory processes. The acquired analogue signal is digitized, amplified, and filtered according to the requirements of the EEG signal of interest. Relevant and precisely defined features of the resulting signal are extracted from the background “noise” generated by billions of active nerve cells. These features are then translated into device commands specific to the BCI-controlled applications\cite{kubler2019history}.

Due to the nature of the acquisition process, the EEG signal can mix with other biological potentials. The greatest influence is on the ocular signal (EOG), which is in the same amplitude range as the EEG (i.e., millivolts) \cite{uriguen2015eeg},\cite{vidal2011analysing}, but it can still be contaminated by muscle (EMG) and cardiac (ECG) signals, both in the millivolt range \cite{reaz2006techniques},\cite{xie2021multi}.
Broadly speaking, artifacts can originate from internal (physiological activities of the subject) and external sources (environmental interference, equipment, electrode pop-up, and cable movement) and contaminate recordings in both temporal and spectral domains \cite{islam2016methods}.  

The signal emitted by the eye movement produces a very strong artifact on EEG signal due to its close proximity to the sensors and abundance of occurrence \cite{woestenburg1983removal}. For this reason eye-blinks are known to substantially contaminate EEG signals, and thereby severely impact the decoding of EEG signals in various medical and scientific applications \cite{agarwal2019blink}. This impact can be measured using Signal to Noise Ratio (SNR). SNR is a dimensionless number that indicates the ratio of the signal power divided  by the noise power contained in a given signal, where the power \(P_s\) is the square of the signal $s(t)$ integrated over time and normalized and $P_N$ is the noise power  \cite{johnson2006signal}. 

\begin{equation}
    SNR =\frac{P_s}{P_N} = \frac{\frac{1}{T} \cdot \int_{0}^{T}s^2(t) \cdot dt }{P_N}
\end{equation}

In the context of detecting eye blink artifacts in EEG waveforms for further removal and signal purification, multiple strategies are proposed in the literature.  Most commonly applied methods require the use of a large number of electrodes and complex equipment for sampling data. Some of these methods are described in \cite{delorme2007enhanced}, where the efficiency of three different Independent Component Analysis (ICA) algorithms were compared in order to detect eye blink occurrences within recorded data, and in \cite{ghosh2018automated}, where Support Vector Machine (SVM) technique was used to identify the blinks, obtaining an accuracy of 98.4\%. Recent methods require fewer EEG channels and simpler equipment, therefore easing and improving the collecting data process. Some studies implementing these recent methods have been reported. \cite{agarwal2019blink} proposed an algorithm able to estimate the timestamps of the start and end of the blinks, \cite{singla2011comparison} compared the efficiency of Support Vector Machine (SVM) against Artificial Neural Network (ANN) on detecting eye opening, closing and blinking, reaching an accuracy of 91.9\% for the SVM and 89.3\% for the ANN in the blink detection experiment, \cite{chambayil2010eeg} created an Artificial Neural Network (ANN) focused on blink detection in EEG signals, obtaining an accuracy of 90.85\%, and \cite{giudice20201d} proposed the use of an one dimensional (1D) convolutional neural network (CNN) with the objective of classify recorded eye blink EEG data between voluntary and involuntary, obtaining an accuracy of 97.92\%.

Although CNN has gained popularity in the last 5 years as the gold standard for image classification, it has been used in many of the state of the art deep learning algorithms \cite{pelletier2019temporal}, these neural networks can also be used for time series classification. Using a CNN for feature extraction has the advantage of not needing a prior filter model or featured engineered treatment as the kernel’s weights are obtained during training and time dependent features will be extracted by the internal structure of convolutional layer \cite{liu2018time}. 

As described by \cite{liu2018time}, CNN can be used for multi-variate or univariate time series, so it can be directly applied for single or multiple channel EEG’s signals. The standard approach is to create a neural network with fixed input shape and one or more stacks
of convolutional layers followed by a pooling layer. The number of kernels and sizes are hyperparameters that can be defined by exploring data from previous experiences with the data format \cite{liu2018time}.

The goal of this work is to create a reliable and user independent algorithm for detecting and removing eye blink in EEG signals using CNN.

\section{Methodology}
The raw EEG data present in the datasets used to train and evaluate this model was collected by Agarwal and Sivakumar \cite{agarwal2019blink} in a series of experiments described in the article “Blink: A Fully Automated Unsupervised Algorithm for Eye-Blink Detection in EEG Signals". Three sets of EEG data collected in those experiments, using the OpenBCI platform with sampling frequency of 250Hz, were used for the purpose of this article. All three sets contain samples obtained while the recruited subjects performed assigned tasks that included blink voluntarily in specific moments, watch a video and read an article. For each participant in the experiments there were generated two files, one with the raw EEG data and the other with the true labels of the instants when eye blinks occurred. The set named EGG-IO contains samples of 20 subjects while blinking voluntarily in specific moments, being the total experiment duration in a range within 75 and 100 seconds. The sets named EEG-VV and EEG-VR contain samples of 12 subjects while reading an article and watching a video respectively, both experiments with a duration of 5 minutes. 

Aiming the highest performance of the convolutional neural network described on this paper, data preprocessing was performed equally on all three sets of data, using the Python programming language. The entire process described below can be checked at \cite{EEG_blink_detector}. Initially, the EEG readings referring to electrodes P1 and P2, along with the time readings, were loaded from the data files into a list of tuples. The same was made with the true labels files, extracting from them the instants when eye blinks were registered. Moving average filter was applied only to the readings from the data files. Based on the information available about the data collection process, tuples with corrupted EEG readings were removed from the list of data. Following the initial steps, two sets were created, the first containing eye blink EEG readings and the second containing EEG readings without eye blinks occurrences. The first set is constituted by same sized arrays taken from the original set of data with arrays of one second windows (two channels with 256) taken from the regions where blink occurrences were registered, the readings that occurred in the exact instants registered on the labels files are the middle point of the windows. Once the windows with blink EEG readings were extracted from the original set, the remaining data was divided in same sized arrays, forming the dataset with no eye blink occurrences. Seeking to balance the quantity of data contained into the datasets, they were resampled. 

Concluding the preprocessing, the datasets were joined together and divided into training, testing and validation sets. The sets were normalized and reshaped following the input format required by the convolutional neural network. Training data from two sets were then combined and fed into the convolutional neural network for training the CNN model.

\section{Results}

The final architecture of the convolutional neural network (CNN) model, used for results analysis over the testing data, was reached by finding the most efficient combination of hyperparameters for each layer. This process was performed by running a random search with the package keras tunner \cite{omalley2019kerastuner}. From the convolutional layer, the hyperparameters selected and tuned were the filters, the kernel size and strides, that correspond respectively to the number of kernels that will be applied over the input data for extracting relevant features, the dimension of these filters and their moving pace. From the dense layers, the only hyperparameters tuned was the number of units, that corresponds to the number of neurons present in each layer. Other important parameters choices for the efficiency of the model were the padding and the activation function. The first one is from the convolutional layer and is responsible for defining an operation mode used to avoid losing information over the input data or shrinking the output data when the convolution filters are applied, the second one is a parameter needed in both convolutional and dense layers and is responsible for assisting the learning process of the CNN.

\begin{figure}
  \centering
  \includegraphics[width=0.98\linewidth, height=9cm]{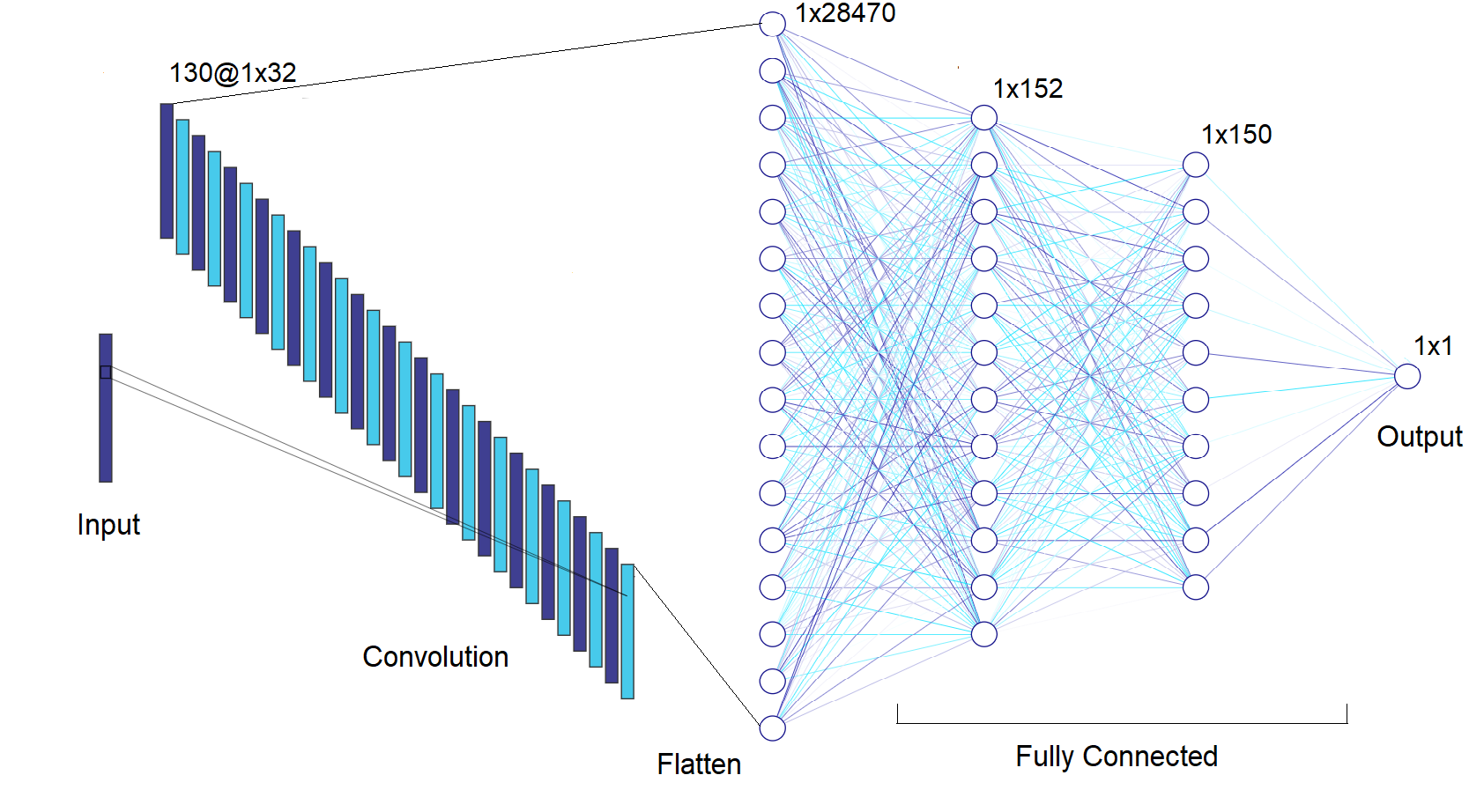}
  \caption{Convolutional neural network build for the study.}
  \label{fig:rede}
\end{figure}

The model used in this paper is represented in Figure \ref{fig:rede} and is constituted by 5 stacked layers. The first layer is a one dimension convolutional layer with 130 filters, a kernel size of 32, the ‘valid‘ padding mode, strides equal to 1 and the rectified linear activation function (ReLU).  The following layer is a flatten layer used to reshape the received data from the previous layer into an one dimensional tensor that can be accepted as input by the next layer. The last three layers are the dense layers that compose the fully connected part of the model. The first dense layer has 152 neurons and uses ReLU activation function, the second dense layer has 150 neurons and also uses ReLU activation function, the last dense layer has only 1 neuron that provides the output of the model, using the sigmoid activation function to discriminate if the received input is an EEG signal of an eye blink occurrence or not.

\begin{figure}[H]
  \centering
  \subfloat[IO set confusion matrix.]{\includegraphics[width=0.45\linewidth,
  height=0.32\linewidth
  ]{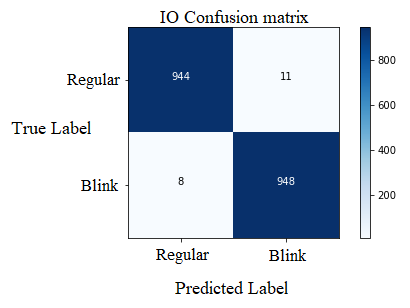}\label{fig:IOcm}}
  \hfill
  \subfloat[VV set confusion matrix.]{\includegraphics[width=0.45\linewidth,
  height=0.32\linewidth
  ]
  {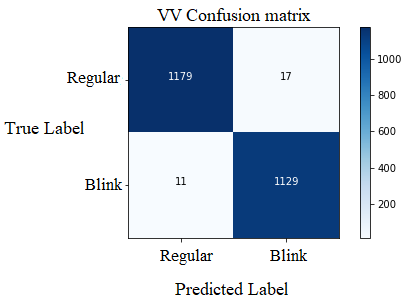}\label{fig:VVcm}}
  \hfill
  \subfloat[VR set confusion matrix.]{\includegraphics[width=0.45\linewidth,
  height=0.32\linewidth
  ]
  {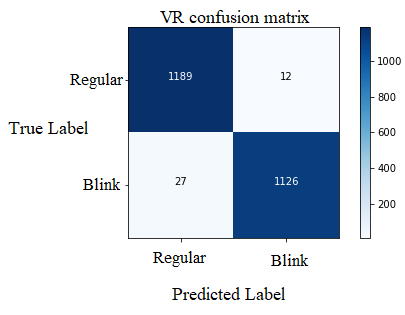}\label{fig:VRcm}}
  \caption{Confusion matrices generated.}
  \label{fig:confusionMatrices}
\end{figure}

The model classification performance over the three sets of validation data was evaluated using four standard metrics named accuracy, precision, recall and f1-score. These values were calculated using the results from the confusion matrix generated for each validation set. These matrices contain the count of true positive, true negative, false positive and false negative classifications.  True positive and true negative correspond to the right classifications made by the model,  while the false positive and false negative correspond to the wrong classifications made by the model. All three matrices were plotted and are represented in Figure \ref{fig:confusionMatrices}, exhibiting a darker blue shade in the true positive and true negative squares, and a lighter blue shade in the false positive and false negative squares. 

Table \ref{table:cnnPerformance} reports the calculated metrics values for each dataset, showing a good and consistent performance in all validation datasets. The fact that each dataset was collected while the participants performed three different tasks demonstrates that the CNN model was able  to learn features and patterns contained in eye blink signals regardless of the situations or the external stimulus present when they were registered. Table \ref{table:literature} provides an overview of the existing methods used in the literature for eye blink classification and their respective accuracies. For comparison reasons, the arithmetic average of the accuracies reached in the three validations sets was calculated, resulting in 98.733\%.

\begin{table}[H]
\caption{CNN performance}
\centering
\begin{tabular}{c c c c c}
\toprule
Validation Dataset & Accuracy & Precision & Recall & F1-Score \\ 
\midrule
IO & 99.057\% & 98.852\% & 99.163\% & 99.007\% \\
VV & 98.801\% & 98.516\% & 99.035\% & 98.775\% \\
VR & 98.343\% & 98.945\% & 97.658\% & 98.297\% \\
 \bottomrule
\end{tabular}
\label{table:cnnPerformance}
\end{table}

\begin{table}[H]
\caption{Literature overview}
\centering
\begin{tabular}{c c c}
\toprule
Study & Method & Accuracy \\
\midrule
Present study & Convolutional neural network (CNN) & 98.733\% \\
\cite{ghosh2018automated} & SVM & 98.4\% \\
\cite{singla2011comparison} & SVM & 91.4\% \\
\cite{singla2011comparison} & Artificial neural network (ANN) & 89.3\% \\
 \bottomrule
\end{tabular}
\label{table:literature}
\end{table}

\begin{figure}[H]
  \centering
  \subfloat[Sample from VR set.]{\includegraphics[width=0.5\textwidth]{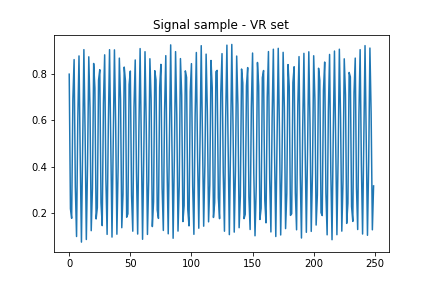}\label{fig:VRruido}}
  \hfill
  \subfloat[Sample from IO set.]{\includegraphics[width=0.5\textwidth]{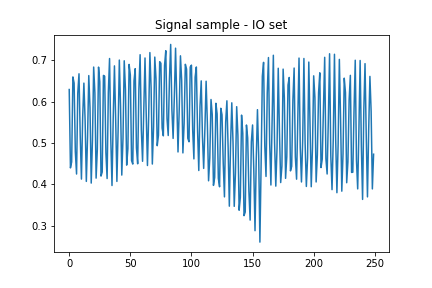}\label{fig:IOruido}}
  \caption{Data contaminated with noise.}
  \label{fig:noiseData}
\end{figure}

Many attempts were made seeking to improve even further the classification performance of the CNN, and, by analyzing the signal samples that were misclassified by the model, a data limitation stood out. Many misclassified samples were contaminated with a substantial amount of noise, this fact hinders and limits classification performance and the learning process of the CNN. Therefore, the improvement of the collected data quality is a key factor to enhance the classification performance of this model. Figure \ref{fig:VRruido} is a signal sample extracted from the IO set and Figure \ref{fig:IOruido} was extracted from the VV set, both signals exemplify the performance barrier mentioned. 

\section{Conclusion}
This work presented a powerful application of the convolutional neural network (CNN) technique for EEG signal artifacts classification. The model used in this study was able to have an embracing understanding of all the features that distinguish a trivial EEG signal from a signal contaminated with eye blink artifacts without being overfitted by specific features that only occurred in the situations when the signals were registered. This aspect is proven by the fact that the model was trained with data from only two of the three sets available, each containing data collected under different external stimulus, and the data from the remaining set was used for validation, having the model achieved an excellent performance on it. Thus, this technique showed its potential to be applied as a tool for removing eye blink artifacts in studies that need real time EEG analyses.


\begin{thebibliography}{10}

\bibitem{stone2013early}
James~L Stone and John~R Hughes.
\newblock Early history of electroencephalography and establishment of the
  american clinical neurophysiology society.
\newblock {\em Journal of Clinical Neurophysiology}, 30(1):28--44, 2013.

\bibitem{kubler2019history}
Andrea K{\"u}bler.
\newblock The history of bci: From a vision for the future to real support for
  personhood in people with locked-in syndrome.
\newblock {\em Neuroethics}, pages 1--18, 2019.

\bibitem{islam2016methods}
Md~Kafiul Islam, Amir Rastegarnia, and Zhi Yang.
\newblock Methods for artifact detection and removal from scalp eeg: A review.
\newblock {\em Neurophysiologie Clinique/Clinical Neurophysiology},
  46(4-5):287--305, 2016.

\bibitem{ramadan2017brain}
Rabie~A Ramadan and Athanasios~V Vasilakos.
\newblock Brain computer interface: control signals review.
\newblock {\em Neurocomputing}, 223:26--44, 2017.

\bibitem{uriguen2015eeg}
Jose~Antonio Urig{\"u}en and Bego{\~n}a Garcia-Zapirain.
\newblock Eeg artifact removal—state-of-the-art and guidelines.
\newblock {\em Journal of neural engineering}, 12(3):031001, 2015.

\bibitem{vidal2011analysing}
M{\'e}lodie Vidal, Andreas Bulling, and Hans Gellersen.
\newblock Analysing eog signal features for the discrimination of eye movements
  with wearable devices.
\newblock In {\em Proceedings of the 1st international workshop on pervasive
  eye tracking \& mobile eye-based interaction}, pages 15--20, 2011.

\bibitem{reaz2006techniques}
Mamun Bin~Ibne Reaz, M~Sazzad Hussain, and Faisal Mohd-Yasin.
\newblock Techniques of emg signal analysis: detection, processing,
  classification and applications.
\newblock {\em Biological procedures online}, 8(1):11--35, 2006.

\bibitem{xie2021multi}
Xiaoyun Xie, Hui Liu, Minglei Shu, Qing Zhu, Anpeng Huang, Xiangpu Kong, and
  Yinglong Wang.
\newblock A multi-stage denoising framework for ambulatory ecg signal based on
  domain knowledge and motion artifact detection.
\newblock {\em Future Generation Computer Systems}, 116:103--116, 2021.

\bibitem{woestenburg1983removal}
JC~Woestenburg, MN~Verbaten, and JL~Slangen.
\newblock The removal of the eye-movement artifact from the eeg by regression
  analysis in the frequency domain.
\newblock {\em Biological psychology}, 16(1-2):127--147, 1983.

\bibitem{agarwal2019blink}
Mohit Agarwal and Raghupathy Sivakumar.
\newblock Blink: A fully automated unsupervised algorithm for eye-blink
  detection in eeg signals.
\newblock In {\em 2019 57th Annual Allerton Conference on Communication,
  Control, and Computing (Allerton)}, pages 1113--1121. IEEE, 2019.

\bibitem{johnson2006signal}
Don~H Johnson.
\newblock Signal-to-noise ratio.
\newblock {\em Scholarpedia}, 1(12):2088, 2006.

\bibitem{delorme2007enhanced}
Arnaud Delorme, Terrence Sejnowski, and Scott Makeig.
\newblock Enhanced detection of artifacts in eeg data using higher-order
  statistics and independent component analysis.
\newblock {\em Neuroimage}, 34(4):1443--1449, 2007.

\bibitem{ghosh2018automated}
Rajdeep Ghosh, Nidul Sinha, and Saroj~Kumar Biswas.
\newblock Automated eye blink artefact removal from eeg using support vector
  machine and autoencoder.
\newblock {\em IET Signal Processing}, 13(2):141--148, 2018.

\bibitem{singla2011comparison}
Rajesh Singla, Brijil Chambayil, Arun Khosla, and Jayashree Santosh.
\newblock Comparison of svm and ann for classification of eye events in eeg.
\newblock {\em Journal of Biomedical Science and Engineering}, 4(1):62, 2011.

\bibitem{chambayil2010eeg}
Brijil Chambayil, Rajesh Singla, and Rameshwar Jha.
\newblock Eeg eye blink classification using neural network.
\newblock In {\em Proceedings of the world congress on engineering}, volume~1,
  pages 2--5, 2010.

\bibitem{giudice20201d}
Michele~Lo Giudice, Giuseppe Varone, Cosimo Ieracitano, Nadia Mammone,
  Arcangelo~Ranieri Bruna, Valeria Tomaselli, and Francesco~Carlo Morabito.
\newblock 1d convolutional neural network approach to classify voluntary eye
  blinks in eeg signals for bci applications.
\newblock In {\em 2020 International Joint Conference on Neural Networks
  (IJCNN)}, pages 1--7. IEEE, 2020.

\bibitem{pelletier2019temporal}
Charlotte Pelletier, Geoffrey~I Webb, and Fran{\c{c}}ois Petitjean.
\newblock Temporal convolutional neural network for the classification of
  satellite image time series.
\newblock {\em Remote Sensing}, 11(5):523, 2019.

\bibitem{liu2018time}
Chien-Liang Liu, Wen-Hoar Hsaio, and Yao-Chung Tu.
\newblock Time series classification with multivariate convolutional neural
  network.
\newblock {\em IEEE Transactions on Industrial Electronics}, 66(6):4788--4797,
  2018.

\bibitem{EEG_blink_detector}
Gustavo Atzingen, Amanda Iaquinta, and Jéssica Toledo.
\newblock Eeg multipurpose eye blink detector using convolutional neural
  network(cnn), April 2021.

\bibitem{omalley2019kerastuner}
Tom O'Malley, Elie Bursztein, James Long, Fran\c{c}ois Chollet, Haifeng Jin,
  Luca Invernizzi, et~al.
\newblock Keras {Tuner}.
\newblock \url{https://github.com/keras-team/keras-tuner}, 2019.

\end{thebibliography}
\end{document}